\documentclass[epj]{svjour}

\usepackage{graphicx}
\usepackage{fancyhdr}
\usepackage{bm}
\def\makeheadbox{{
 }}
\def \sle   {\ensuremath{ \tilde{\ell}              }}

\def \neu   {\ensuremath{ \mathrm{\tilde{\chi}^0}           }}
\def \neue  {\ensuremath{ \mathrm{\tilde{\chi}^0_1}         }}
\def \neuz  {\ensuremath{ \mathrm{\tilde{\chi}^0_2}         }}

\def \cha   {\ensuremath{ \mathrm{\tilde{\chi}^{\pm}}    }}

\def \lep   {\ensuremath{ \ell                      }}

\def \msle   {\ensuremath{ m_{\sle}                }}

\def \mneuz  {\ensuremath{ m_{\neuz}               }}

\newcommand{\et}{\ensuremath{E_\mathrm{T}                        }}
\newcommand{\etmiss}{\ensuremath{\not\!\!\et                     }}
\newcommand{\sigbr}{\ensuremath{\sigma\times\mathrm{BR}(3\ell)   }}
\newcommand{\ptthree}{\ensuremath{p_T^{\ell3}}}
\newcommand{\dzero}{D\O\                                          }

\def\mytitle{My title} 
\def\myauthors{My name}  
\def\mytype{My type of session}
\def\mysession{My session}

\def\mytitle{Search for Supersymmetry in Trilepton Final States with the D\O\ detector} 
\def\myauthors{Olav Mundal}    
\def\mytype{Contributed Talk}    
\def\mysession{Colliders - SUSY Phenomenology}

\begin{document}
\title{Search for Supersymmetry in Trilepton Final States with the D\O\ Detector.}

\author{Olav Mundal for the D\O\ Collaboration
}                     
%
%
\institute{University of Bonn, Nussallee 12, 53115, Bonn, Germany}
%
\date{}
\abstract{
Data taken by the D\O\ experiment at the proton-antiproton collider
at Tevatron has been analyzed to search for signatures consistent 
with decay of Charginos and Neutralinos. The search is performed
in final states with three leptons and missing transverse energy.
No excess above the Standard Model expectation is observed and 
limits on the production cross section times Branching fraction
are set.
} 
\maketitle
\section{Introduction}
\label{intro}

The proton-antiproton collider Tevatron delivers at present
the world's highest center of mass energies with
$\sqrt{s}=1.96$ TeV. The D\O\ experiment, one of the two
experiments at the Tevatron, has collected data corresponding
to an integrated luminosity of close to 3.0 fb$^{-1}$ 
in RunIIa and RunIIb. Here one analysis based on the 
RunIIb dataset and four analyses based on 
the RunIIa dataset will be described.
The search for Supersymmetry (SUSY) is one of the main
goals at D\O\ . The Charginos ($\tilde{\chi}^{\pm}$) and Neutralinos ($\tilde{\chi}^0$),
the superpartners of the Standard Model (SM) gauge and Higgs bosons, 
are of particular interest because
the decay mode into three charged leptons and missing
transverse energy provides a clean signature with small
SM background expectations.

\section{Supersymmetric model and decay}
\label{sec:1} 

\begin{figure}
\begin{center}

\includegraphics[width=0.35\textwidth,height=0.25\textwidth,angle=0]{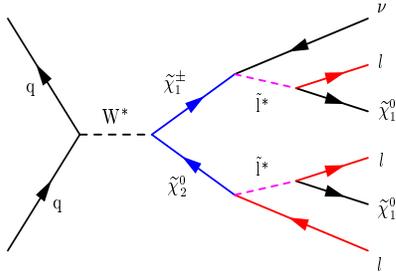}
\caption{An example of production and decay modes for the signal points considered in the analyses.}
\label{fig:1}     

\end{center}
\end{figure}

Supersymmetry postulates a symmetry between
bosons (integer spin) and fermions (half integer spin). 
For every particle in nature there is 
a supersymmetric partner that differs in spin
with $\frac{1}{2}$. No evidence has so 
far been found for the existence of SUSY particles
and limits have been set by previous
experiments ~\cite{lep_susy}.
The analyses presented here describe a 
search for the SUSY partners of the charged and 
neutral electroweak gauge and Higgs bosons.
The search is performed in final
states with two identified leptons ($e$ or $\mu$), 
a third track and large
missing transverse energy. The analysis is 
based on the Minimal Supersymmetric extension
of the Standard Model (MSSM) with R-parity conservation ~\cite{mssm}.
R is given by $(-1)^{3(B-L)-2s}$. R parity conservation leads
to a stable Lightest Supersymmetric Particle (LSP), in this 
case the lightest Neutralino. 
More specifically, the results are interpreted using minimal 
supergravity (mSUGRA) models. mSUGRA models are characterized
by five parameters: $m_{0}$, the common fermion mass at GUT scale,
$m_{1/2}$, the common scalar mass at GUT scale,
$A_{0}$, the trilinear coupling, 
tan$\beta$, the ratio of vacuum expectation values 
of the two Higgs fields
and sign($\mu$), the Higgs(ino) mass parameter. 
At the Tevatron, the dominant channel for producing Charginos 
and Neutralinos is the s-channel via an 
off-shell W boson ~\cite{susyhad}.
The t-channel production takes place via squark exchange and
this diagram interferes destructively with the s-channel. 
The t-channel is suppressed at higher squark masses.
See Fig.~\ref{fig:1} for an example of production and
decay of Charginos and Neutralinos. 
The branching fraction of Charginos and Neutralinos into
leptons depends on the slepton mass. At high slepton mass
$W/Z$ exchange is dominant which leads to small branching
fraction into leptons (large $m_{0}$ scenario). The leptonic
branching fraction for 3-body decays is at maximum when
$\msle$ is just above $\mneuz$ ($3l$-max scenario).
The result is interpreted in a scenario without slepton
mixing and the masses of the right handed sleptons are assumed 
degenerated.

\begin{table}
\caption{SUSY parameters for three reference signal points ordered
by decreasing masses. All points have
tan$\beta$ = 3, $A_{0}=0$, $\mu>0$. All masses are in GeV.}
\label{tab:1}       
\begin{tabular}{lllll}
\hline
Point & $m_{\cha}$  & $m_{\neuz}$ & $m_{\neue}$ &  \sigbr   \\

\hline
\noalign{\smallskip}
High mass   & 150 & 152 & 82 & 0.0366 \\
Medium Mass & 125 & 127 & 69 & 0.123\\
Low mass    & 115 & 118 & 63 & 0.1984\\
\noalign{\smallskip}\hline
\end{tabular}
\vspace*{1cm}  
\end{table}

\section{Search strategy} \label{sec:3}

\subsection{General}

The trilepton final states are 
characterized by three 
charged leptons and 
missing transverse energy, ($\etmiss$), 
due to the presence of neutrinos and
neutralinos in the final state.
One of the leptons in the final state has in general
very low $p_{T}$. To increase the efficiency of the 
selection, only two fully identified 
leptons ($e$ or $\mu$) are required while a high quality, 
isolated track is required 
instead of the third lepton. 
This approach keeps the
efficiency high and is sensitive to 
tracks from electrons, muons and 
taus.\\
In some parts of the parameter space, 
however, this strategy breaks down.
When $\msle$ is just below $\mneuz$, the third lepton 
is extremely soft and the trilepton
analyses lose their sensitivity. Instead of requiring
two charged leptons and an isolated track, a pair of like-sign leptons
of the same flavour (muons) is required.
This is enough to suppress the SM 
background to an acceptable level.
Signal parameter combinations have been generated with tan$\beta$=3 and
Chargino masses in the range of 115-150 GeV
using the Les Houches Accords (LHA) ~\cite{lha}. See Table ~\ref{tab:1}.
The main background components are vector boson pair production,
($WW$, $WZ$, $ZZ$) and $Z/\gamma^{*}$ and $W$ production.  
In the case of the Like Sign muon analysis, QCD multijet production
is the most important background.
All background processes are generated with PYTHIA ~\cite{pythia}, except 
QCD multijet background which is taken directly from data
by inverting lepton identification variables.
The main focus in the following will be on
the final state with two identified electrons and a third track.
This analysis was newly updated with RunIIb data. Three other
RunIIa analyses with final states with two muons 
and a third track, one electron,
one muon and a third track and finally 
the Like Sign muon analysis are 
also discussed in some detail.  
In the following the following notation 
will be used to label the different analyses:
$ee+\lep$ for the final state with two
electrons and a third track,
$e\mu+\lep$ for the final state
with one electron, one muon and a third
track and $\mu\mu+\lep$ for the final
state with two muons and a third
track.

\begin{center}
\begin{figure}
\includegraphics[width=0.45\textwidth,height=0.45\textwidth,angle=0]{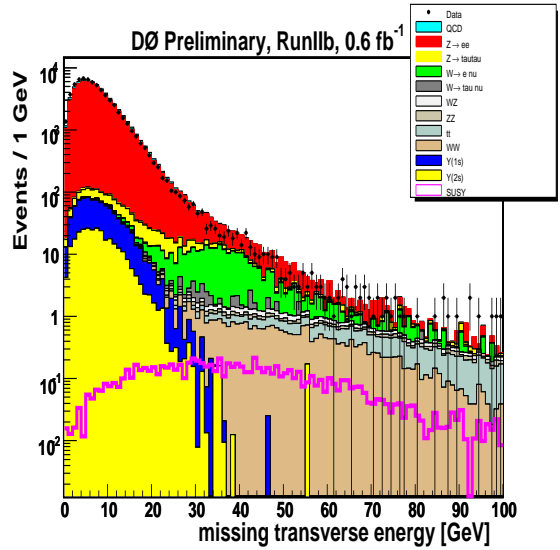}
\caption{Distribution of the missing transverse energy, $\etmiss$, at preselection level in the 
$ee+\lep$ analysis based on RunIIb data.}
\label{fig:2}    
\end{figure}
\end{center}

\subsection{$ee+\lep$ selection} \label{eel}

\begin{center}
\begin{figure}
\includegraphics[width=0.35\textwidth,height=0.25\textwidth,angle=0]{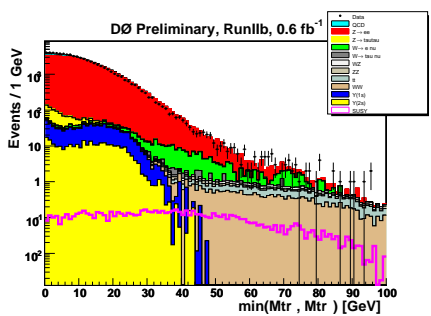}
\caption{The distribution of the minimum transverse mass at preselection level.}
\label{fig:3}     
\end{figure}
\end{center}

\begin{center}
\begin{figure}
\includegraphics[width=0.35\textwidth,height=0.25\textwidth,angle=0]{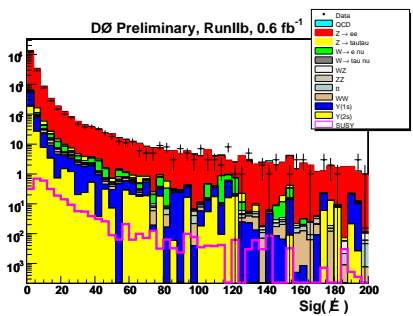}
\caption{The distribution of the scaled missing transverse energy, Sig($\etmiss$), at preselection level.}
\label{fig:4}      
\end{figure}
\end{center}

\begin{center}
\begin{figure}
\includegraphics[width=0.35\textwidth,height=0.25\textwidth,angle=0]{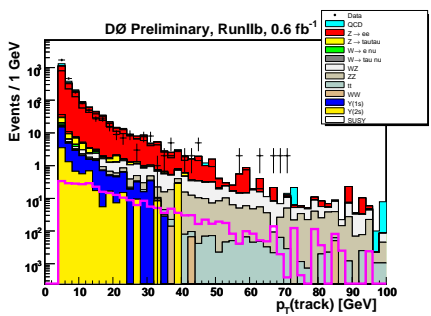}
\caption{The isolated track transverse momentum, \ptthree, at preselection level.}
\label{fig:5}      
\end{figure}
\end{center}

The selection requires two electrons with $p_{T}>8, 12$ GeV. The electrons
have to be isolated and to match a reconstructed track in $\eta$ and $\phi$.
Both electrons must stem from the primary vertex and events 
with both electrons in detector regions with poor resolution are
rejected. To supress the large contribution
from $Z \rightarrow ee$ events, several cuts are made. The invariant mass
of the two electrons is required to be less than 60 GeV and above 18
GeV. Furthermore, the opening angle in the transverse plane has to be
below 2.9 rad, $\Delta \phi < 2.9$. Events with large jet activity
are rejected to reduce the contribution from $t\bar{t}$.
$Z/\gamma^{*}$ and QCD events have small values of $\etmiss$ and an
important cut to reject these backgrounds is to require $\etmiss$ 
to be greater than 22 GeV. 
See Fig.~\ref{fig:2} for the $\etmiss$ distribution at preselection 
level.
Mismeasurements of the lepton energies lead to small values
of the quantity transverse mass. 
Transverse mass is defined as
$m_{T} = \sqrt{p_{T} \etmiss(1-\Delta \phi(e,\etmiss ))}$ and is 
required to be greater than 20 GeV for both electrons. 
Larger values of $\etmiss$ can also be produced due to 
fluctuations of the jet energy depoistions. To address this 
fact, the quantity Sig($\etmiss$) is required to be in excess of 8.
Sig($\etmiss$) is defined
by dividing the \etmiss\ by
the jet resolution projected onto the 
direction of the \etmiss\, 
$\sigma_{E_T^j\!\parallel\,,\etmiss}$.
See Fig.~\ref{fig:3} and Fig.~\ref{fig:4} for the
distributions of $m_{T}$ and Sig($\etmiss$) at preselection level.\\
The third track is required to come from the same vertex as the two
electrons but also to be well separated from them 
($\Delta R = \sqrt{(\Delta \eta)^{2}+(\Delta \phi)^{2}}>$0.4).
The tracks are also required to have a $p_{T}$ greater than 4. 
Isolation both in calorimeter and tracker is required. 
See Fig.~\ref{fig:5} for the $p_{T}$ distribution of the 
leading isolated track in events where an isolated track is found.
The final cut in the selection is to require the 
product of track $p_{T}$ and $\etmiss$ to be greater
than 220 GeV$^{2}$. See Table ~\ref{tab:2} 
for expected number 
of signal and background events as well as
observed events.

\subsection{Likesign $\mu\mu$ selection} \label{lsmu}

The preselection in the like sign di-muon channel requires
two muons of the same charge with transverse momenta $p_{T}>$ 5 GeV which
are not back to back ($\Delta \phi < $2.9). Three muons events with
opposite sign pairs of invariant mass above 65 GeV are discarded.
As in the other analyses a set of cuts related to \etmiss\
are applied. The missing transverse energy must be in excess of 10 GeV 
while the transverse mass is required to be in the range between 15 GeV
and 65 GeV. The significance of \etmiss\ must be greater than
12. See Table ~\ref{tab:2} for signal and background expectation as well as 
events observed in data. A more detailed description can be found in 
~\cite{vincent}.

\subsection{$e\mu+\lep$ and $\mu\mu+\lep$ selection} \label{emu}

A detailed description of the $e\mu+\lep$ and $\mu\mu+\lep$ analyses 
can be found in ~\cite{winter07}. 
In the following a brief summary of the two selections will be given.
For both analyses, the requirement of a third, isolated track is the
same as in the $ee+\lep$ discussed above. 
In the $e\mu+\lep$ analysis the events selected have at least one 
electron with $p_{T}>$ 12 GeV and one muon with $p_{T} > $  8 GeV
and the invariant mass of the electron and the muon must be
greater than 15 GeV and smaller than 100 GeV. 
Both the electron and the muon must stem
from the primary vertex.
To remove potential WZ background, events 
with either two muons or two electrons are removed if the their
invariant mass is larger than 70 GeV.
The \etmiss\ is required to be in excess of 10 GeV and Sig($\etmiss$)
must be above 8 GeV. The transverse mass with \etmiss\
must be above 20 GeV and below 90 GeV for 
both the electron and the muon. The transverse mass
of the third track and \etmiss\ must be greater than 8 GeV and
the invariant mass between either the electron or muon and the 
third track must be below 70 GeV.
The $\mu\mu+\lep$ selection requires two isolated muons stemming
from the primary vertex with $p_{T}>$ 12 and 8 GeV. The 
invariant mass is required to be at least 24 GeV and not greater
than 60 GeV and the angle between the two muons in the 
transverse plane, $\Delta \phi$, has to be below 2.9 radians.
Again several cuts related to \etmiss\ are applied: \etmiss\ $>$ 
20 GeV, Sig($\etmiss$) $>$ 8 and the transverse mass is 
required to be above 20 for both muons and \etmiss.
For the track, the transverse mass with \etmiss\ must be above
8 GeV and events where the invariant mass 
between the leading muon and \etmiss\ is above 80 GeV are 
rejected.
A summary of expected number of signal and background events
as well as observed events can be found in Table ~\ref{tab:2}.

\section{Results} \label{results}

\begin{center}
\begin{figure}


\includegraphics[width=0.35\textwidth,height=0.25\textwidth,angle=0]{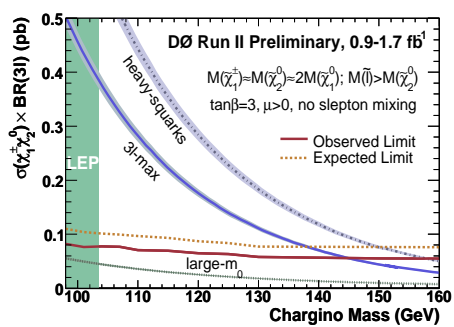}

\caption{Limit on \sigbr\ as a function of
$\cha_1$ mass, in comparison with
the expectation for several SUSY scenarios. The red line corresponds to the observed limit.
PDF and renormalization/factorization scale uncertainties are shown as shaded bands. }
\label{fig:6}       
\end{figure}
\end{center}

\begin{center}
\begin{figure}


\includegraphics[width=0.35\textwidth,height=0.25\textwidth,angle=0]{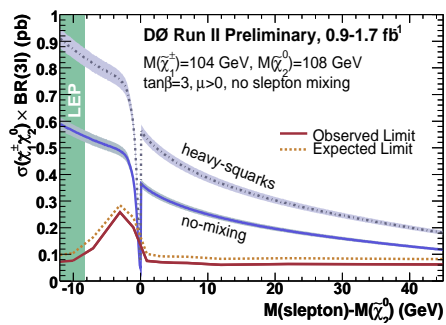}

\caption{Limit on \sigbr\ as a function of the mass difference between sleptons and $\neu_2$,
in comparison with the expectation for the MSSM (no mixing) and the heavy-squarks scenario (see text). PDF and
renormalization/factorization scale uncertainties are shown as shaded bands. }
\label{fig:7}       
\end{figure}
\end{center}

The numbers of observed candidate events as well as the expectation
for signal and background are shown in Table ~\ref{tab:2}. No evidence for
the production of Charginos and Neutralinos is observed and therefore 
an upper limit on the product of cross section and 
leptonic branching fraction, \sigbr, is set. To avoid double counting
between the analyses, events found by more than one
analysis are assigned to the analysis with best signal to
background ratio and removed from all others.
Systematic uncertainties are small compared
to statistical uncertainties due to limited
MC statistics.
The expected and observed limit range between 0.75 and 0.11 pb.
Assuming the $3\lep$-max scenario, see Section ~\ref{sec:1},
the cross section limit can be
translated into a mass limit of 140 GeV. 
See Fig.~\ref{fig:6} for the limit of \sigbr as a function
of the Chargino mass. 
If sleptons are light so that two body decay of 
Charginos and Neutralinos can take place
and the mass difference between 
the sleptons and next to lightest
neutralino is small, the Like Sign muon analysis remains
the only sensitive because of the low $p_{T}$ of the third lepton.
This leads to higher limits in this region of parameter space.
See Fig.~\ref{fig:7} for the limit of \sigbr as a function
of the mass difference between sleptons and next to lightest
neutralino.\\

\begin{table}
\caption{ Number of events expected and observed in data and signal predictions for
the different final states. The integrated luminosity is also given.}
\label{tab:2}       
\begin{tabular}{lllll}
\hline
\noalign{\smallskip}
Final state          & ${\cal L}_{int}$  & Background    & Data  &Signal   \\ 
                      & [fb$^{-1}$]\\ \hline
$ee+\lep$             & 0.6                    & $1.0\pm0.3$               & 0     & 0.5-2.1 \\
$ee+\lep$             & 1.1                    & $0.76\pm0.67$             & 0     & 1.0-5.3 \\
$\mu\mu+\lep$         & 1.0                    & $0.32\pm^{0.78}_{0.08}$   & 2     & 1.5-1.8 \\
$e\mu+\lep$           & 1.1                    & $0.94\pm^{0.40}_{0.18}$   & 0     & 2.0-2.6 \\
$\mu^{\pm}\mu^{\pm}$  & 0.9                    & $1.1\pm0.4$               & 1     & 0.5-2.0 \\
\noalign{\smallskip}\hline

\end{tabular}

\vspace*{1cm}  
\end{table}

%

\section{Conclusion and Outlook} \label{concl}

A search for Charginos and Neutralinos has been performed in
final states with three charged leptons and missing 
transverse energy. No evidence for SUSY has been found and
limits on \sigbr\ have been set. With prospects of
up to 8 fb$^{-1}$ of integrated luminosity the 
limits will continue to improve. More difficult
regions of phasespace will also be probed. For high
Chargino masses, the decay of Charginos and Neutralinos
can go via an on shell $Z$ boson, making the search
more challenging since the current mass cuts have
to be changed and the background from $Z$ bosons
will increase significantly.


\begin{thebibliography}{999}
%
%
\bibitem{lep_susy}
  LEPSUSYWG, ALEPH, DELPHI, L3 and OPAL experiments,note LEPSUSYWG/01-07.1,
  ({\tt http://lepsusy.web.cern.ch/lepsusy} {\tt /Welcome.html})


\bibitem{mssm}
  H.P. Nilles,
  Phys. Rep. \textbf{110} (1984) 1;\\
  H.E. Haber and G.L. Kane,
  Phys. Rep. \textbf{117} (1985) 75.
\bibitem{susyhad}
  W.Beenakker {\it et al.,}
  {\it The production of Charginos/Neutralinos and Sleptons at Hadron Colliders},
  hep-ph/9906298.
\bibitem{lha}
         P. Skands {\it et al.,} JHEP 07, 036 (2004).
\bibitem{pythia}
        T. Sjostrand,
        Comp. Phys. Commun. \textbf{82} (1994) 74,
        \mbox{\texttt{CERN-TH 7112/93} (1993)}.
\bibitem{vincent}
         V. Abazov {\it et al.,} (D\O\ Collaboration),
        {\it Search for the associated production of charginos and neutralinos in like sign dimuon channel},
        \dzero Conference Note 5126.
\bibitem{winter07}
         V. Abazov {\it et al.,} (D\O\ Collaboration),
        {\it Search for the Associated Production of Chargino and Neutralino in Final States with Three Leptons},
        \dzero Conference Note 5348.

\end{thebibliography}
\end{document}